\documentclass[12pt]{iopart}

\usepackage{epsfig}
\begin{document}

\title[Min-oscillations in E.~coli]{Min-oscillations in \textit{Escherichia coli} induced by interactions of membrane-bound proteins}

\author{
Giovanni Meacci and Karsten Kruse
\footnote[3]{To
whom correspondence should be addressed (karsten@mpipks-dresden.mpg.de)}
}

\address{
Max-Planck Institut f\"ur Physik komplexer Systeme, 
N\"othnitzerstr. 38, 01187 Dresden, Germany
}

\begin{abstract}
During division it is of primary importance for a cell to correctly
determine the site of cleavage. 
The bacterium \textit{Escherichia coli} 
divides in the center, producing two daughter cells of equal size.
Selection of the center as the correct division site is in part 
achieved by the Min-proteins. 
They oscillate between the two cell poles and thereby 
prevent division at these locations. Here, a  phenomenological description for these oscillations is presented, where lateral interactions between proteins on the cell membrane play a key role. 
Solutions to the dynamic equations are compared
to experimental findings.
In particular, the temporal period of the oscillations is measured as a function of the cell length and found to be compatible with the
theoretical prediction.
\end{abstract}


\pacs{87.16.Ac, 87.17.Ee, 82.39.Rt}

\maketitle

\section{Introduction}

During division a cell has to specify in particular the location of cleavage. 
In the rod-shaped bacterium \textit{Escherichia coli}, the division plane is 
determined by the location of the Z-ring~\cite{lutk02}. 
This structure is built from FtsZ-filaments and forms on the inner bacterial membrane. 
The position of the Z-ring in turn is first of all determined by the distribution of 
the nuclear material inside the cell. A mechanism termed ``nucleoid occlusion'' restricts 
the formation of the ring to regions void of DNA~\cite{YM99,wu04}. After duplication and 
segregation of the chromosome, three locations of possible ring formation remain: at the 
cell center and close to the two cell poles. Selection of the center as the correct 
division site is achieved by the Min system~\cite{deBCR89,BL93}. 
Deletion of any of the Min proteins results in division septa forming
close to one of the two cell poles in about 50\% of all divisions. In
these cases, DNA-free mini-cells are formed~\cite{adle67}.

The Min system consists of three proteins, MinC, MinD, and MinE. 
Out of these, MinC induces the depolymerization of FtsZ-filaments and inhibits the formation of the Z-ring~\cite{HMPL99}. The distribution of MinC on the membrane changes periodically with time such that in one half of the cycle, MinC accumulates at one pole while it accumulates at the opposite pole in the second half of the cycle~\cite{HL99,RdeB99a}.  Formation of the Z-ring is thereby
suppressed at the cell poles. The temporal period of the oscillation ranges between 40s and 120s in wild-type cells. In bacteria of a length that exceeds a certain threshold, a striped oscillatory pattern appears, where the number of stripes increases with increasing cell length. This observation is indicative of an intrinsic spatial wave-length of the oscillations.

The oscillations of MinC require the presence of both MinD and MinE, which themselves also oscillate~\cite{RdeB99,HMdeB01}. In fact, MinC binds to MinD and follows its dynamics~\cite{RdeB99}.  
In contrast, MinE is mostly localized in a ring structure which oscillates 
around the center of the bacterium. Remarkably, MinC is not 
necessary to generate oscillations, as MinD and MinE oscillate also in
 the absence of MinC.  The behavior of MinD and MinE has been 
 characterized by intensive biochemical and genetic studies. \textit{In 
 vitro} experiments have shown that the ATPase MinD has a high 
 affinity for the inner bacterial membrane if ATP is 
 present~\cite{HGL02}. For concentrations of MinD exceeding a critical 
 value, filamentous MinD aggregates are formed on the 
 membrane~\cite{HGL02,SVR02}. The formation of MinD aggregates
  is likely to be a two step-process, where MinD first binds to the 
  membrane and then self-assembles~\cite{HGL02}. Indeed, in 
  presence of ATP$\gamma$S, a non-hydrolysable analog of ATP, MinD associated with the membrane but failed to form filaments. As for 
  MinE, it associates with the membrane only in the presence of MinD. 
  There, it stimulates hydrolysis of the ATP bound to MinD, which 
  eventually drives the proteins off the membrane~\cite{HGL02}. 
  
These \textit{in vitro} results are compatible with the behavior of MinD and 
  MinE \textit{in vivo}. In MinD depleted cells, it was observed that 
  MinE is dispersed in the cytosol, while MinD is homogenously 
  distributed on the cytosplasmic membrane if MinE is 
  absent~\cite{RdeB99}. Furthermore, helical MinD aggregates have been observed to form on the inner membrane~\cite{SLR03}. The significance of the helical structures for the oscillation mechanism is still not understood. Finally, the oscillations do not depend on the synthesis and degradation of the Min-proteins~\cite{RdeB99}.

Theoretical investigations of the Min-system suggest that the periodic translocations of the Min proteins can be attributed to a collective effect of many interacting molecules resulting from a dynamic instability~\cite{MdeB01,HRdeV01,K02,HMW03}. Central to all proposed mechanisms is the attachment of MinD to the cytoplasmic membrane, recruitment of MinE to the membrane by MinD, and dissociation of MinD from the membrane induced by MinE. The mechanism proposed by Meinhardt and deBoer~\cite{MdeB01}  belongs to the class of classical reaction-diffusion systems with short-range activation and long-range inhibition. The synthesis and degradation of the Min proteins play an essential role. Howard and colleagues~\cite{HRdeV01} assume that MinD and MinE form complexes in the cytoplasm, which then bind to the membrane. Membrane-binding is hampered by MinE present on the membrane. 
Furthermore, 
the protein number is conserved. The same holds 
for the mechanism presented in Ref.~\cite{K02}. There, however, first MinD binds to the membrane and then recruits MinE. More importantly, aggregation of membrane-bound MinD is essential. In contrast to reaction-diffusion systems, the instability is here driven by the aggregation current of MinD. 
Rather
similar to this mechanism is the one proposed by Huang et al.~\cite{HMW03} more recently. In contrast to Ref.~\cite{K02}, aggregation is there assumed to be a consequence of MinD binding cooperatively to the membrane. 
This seemingly small difference in the formation of membrane-bound
MinD aggregates has remarkable consequences.
Firstly, it is essential to describe the Min dynamics in a three-dimensional geometry. Secondly, a finite ADP to ATP exchange rate for cytosolic MinD is a key ingredient. 
As transport is purely diffusive, the instability leading to the oscillations
is in this case essentially of  the same kind as in the other reaction diffusion systems~\cite{MdeB01,HRdeV01}.

In this work we re-investigate the mechanism proposed in 
Ref.~\cite{K02}. There, the aggregation of membrane-bound MinD
was formulated in terms of a kinetic hopping model. Here, we will 
use a phenomenological description, 
which allows for a quantitative comparison with experimental results.
The paper is organized as follows. First, we will describe the equations governing the dynamics of the protein distributions in the cytosol and on the membrane. We then analyze the system in the limiting case of homogenous cytosolic protein distributions and discuss the oscillatory solutions. The dependence of the temporal oscillation period on the system length is compared to experimental data. Afterwards we discuss possible mechanisms underlying the formation of the MinE-ring. Finally, we discuss our results in relation to the other proposed mechanisms as well as 
implications for 
possible future
experiments.

\section{Dynamic equations}

\begin{figure}
\centering
\includegraphics*[width=12 truecm, angle=0]{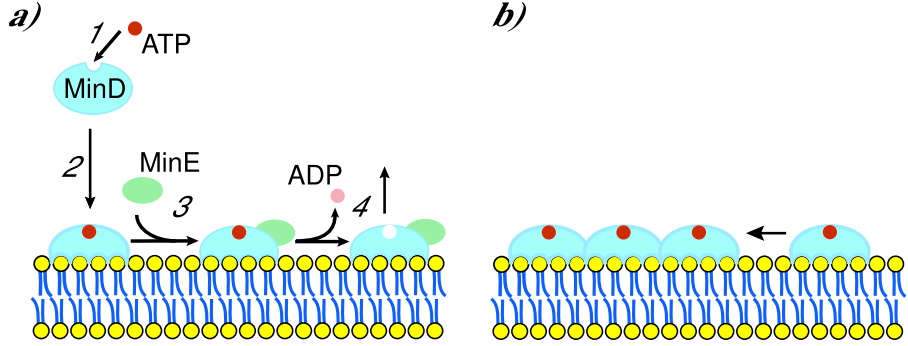}
\caption{
Schematic representation of the dynamics of MinD and MinE. a) Exchange of MinD and MinE between the cytosol and the membrane. 1) MinD undergoes a conformational change upon binding ATP, 2) ATP-bound MinD binds to the membrane, 3) MinE binds to membrane-bound MinD, and 4) MinE-induced ATP hydrolysis leads to detachment of MinDE-complexes from the membrane. b) Interaction of membrane-bound proteins leads to the formation of MinD aggregates.
}
\label{fig:schema}
\end{figure}

As mentioned above, the periodic changes in the distributions of the Min proteins require the presence of MinD and MinE, but not of MinC. Therefore, we will focus in the following on the dynamics of MinD and MinE. Motivated by the observations reported above, the dynamics of the Min proteins is assumed to be driven by four properties of the Min proteins~\cite{K02}: i)  a high affinity of ATP-bound MinD for the membrane, ii) a high affinity of MinE for membrane-bound MinD, iii) a MinE-induced increase of the ATP hydrolysis-rate by MinD, which leads to the detachment of MinDE-complexes from the membrane, and iv) interactions between membrane-bound proteins. The last property accounts for the formation of MinD aggregates on the membrane, which is likely to result from self-assembly of membrane-bound MinD~\cite{HGL02}. In addition, proteins are transported by diffusion. A schematic representation of the Min dynamics is given in Fig.~\ref{fig:schema}. 

Formally, the dynamics is given in terms of the concentrations of cytosolic MinD and MinE, $c_D$ and $c_E$\footnote{MinE forms dimers~\cite{king99} and $c_{E}$ is actually the distribution of MinE dimers. In the following, the term ``MinE molecules'' refers to these dimers.}, as well as the concentrations of membrane-bound MinD and MinDE-complexes, $c_d$ and $c_{de}$. In the direction perpendicular to the long axis of the bacterium, diffusion homogenizes the cytosolic distributions on time scales that are short as compared to the temporal oscillation period. Assuming in addition that MinD aggregates into a linear structure on the membrane, the dynamical equations for the protein densities in the cell can thus be reduced such that they depend only on the position $x$ along the long axis of the bacterium, see~\ref{app:reduction}.  Explicitly, 
\begin{eqnarray}
\label{eq:dynD}
\partial_tc_D & = & -\omega_D(c_{\rm max}-c_d-c_{de})c_D+\omega_{de}c_{de} + D_D\partial_x^2 c_D\\
\label{eq:dynE}
\partial_tc_E & = & \omega_{de}c_{de}-\omega_Ec_dc_E + D_E\partial_x^2c_E\\
\label{eq:dynd}
\partial_tc_d & = & \omega_D(c_{\rm max}-c_d-c_{de})c_D - \omega_Ec_dc_E  - \partial_xj_d\\
\label{eq:dynde}
\partial_tc_{de} & = & -\omega_{de}c_{de} + \omega_Ec_dc_E 
-\partial_{x} j_{de}
\end{eqnarray}
The properties (i)-(iii) lead to an exchange of MinD and MinE between the cytosol and the membrane. The corresponding reactions are described as first and second order processes. The density of available binding sites for MinD on the membrane is given by $c_{\rm max}-c_d-c_{de}$, where $c_{\rm max}$ is the maximal possible value for the protein density on the membrane, and  $\omega_D(c_{\rm max}-c_d-c_{de})$ is the binding rate of MinD to the membrane.
The binding rate of MinE to membrane-bound MinD is $\omega_E c_{d}$, while $\omega_{de}$ is the detachment rate of MinDE complexes from the membrane. We assume the complexes to consist of one MinD and one MinE molecule. $D_D$ and $D_E$ are the respective diffusion constants for cytosolic MinD and MinE, and the interactions of membrane-bound proteins are captured by the currents $j_d$ and $j_{de}$. Note, that in these equations the rebinding of ATP to MinD after detachment from the membrane is assumed to occur on a sufficiently short time-scale such that it does not need to be incorporated explicitly. The effect of a finite ATP exchange rate will be discussed below.  

The current of membrane-bound MinD has a diffusive part and a part due to the interaction between MinD proteins. In order to capture generic effects of the interaction, the current of membrane-bound MinD is taken to be of the Cahn-Hilliard form. Explicitly,
\begin{equation}
j_{d} = -D_d\partial_{x}c_{d}+c_d(c_{\rm max}-c_d-c_{de})[k_1\partial_xc_d
+k_2\partial_x^3c_d+\bar k_{1}\partial_{x}c_{de}+\bar k_{2}\partial_{x}^{3}c_{de}].\quad
\end{equation}
In this expression, $D_{d}$ is the diffusion constant of the MinD proteins on the membrane and the coefficients $k_{1}$ and $k_{2}$ are phenomenological parameters that describe the interaction between MinD molecules. Possible modifications of this interaction
due to the presence of MinE are taken into account by the parameters $\bar k_{1}$ and $\bar k_{2}$ that describe the interaction between MinD and MinDE-complexes.  Note, that for an attractive interaction $k_{1}>0$, while $k_{1}<0$ in the opposite case. Stability on small length scales requires $k_{2}\ge0$. The current of MinDE complexes has the same form, but for simplicity will be omitted in the following. 

Finally, the boundary conditions have to be specified. We impose zero flux at the boundaries, such that the total protein numbers
\begin{eqnarray}
\label{eq:consD}
\int_{-L/2}^{L/2} dx\;(c_D+c_d+c_{de}) & \equiv & L{\cal D}\\
\label{eq:consE}
\int_{-L/2}^{L/2} dx\;(c_E+c_{de}) & \equiv & L{\cal E}
\end{eqnarray}
are conserved. Here, $L$ denotes the length of the system and $L{\cal D}$ and $L{\cal E}$ are the total numbers of MinD and MinE molecules in the system, respectively.

\section{Homogenous cytosolic distributions}

We now analyse the dynamic equations~(\ref{eq:dynD})-(\ref{eq:dynde}) in the limiting case of homogenous  cytosolic MinD and MinE distributions, i.e., $c_{D}(x,t)=c_{D}(t)$ and $c_{E}(x,t)=c_{E}(t)$. This corresponds to the case where the times needed for MinD and MinE to diffuse along the whole length of the bacterium, $L^2/D_D$ and $L^2/D_E$, respectively, are short as compared to all other relevant time-scales involved. In this case, the dynamics of the cytosolic distributions is described by ordinary differential equations
\begin{eqnarray}
\frac{d}{dt}c_D & = & -\omega_D(c_{\rm max}-{\cal D}+c_D)c_D + \omega_{de}({\cal E} -c_E)\\
\frac{d}{dt}c_E & = & -\omega_E({\cal D}-{\cal E} -c_D+c_E)c_E + \omega_{de}({\cal E} - c_E)\quad.
\end{eqnarray} 
Here, the distributions of membrane-bound MinD and MinDE have been eliminated using equations (\ref{eq:consD}) and (\ref{eq:consE}). 

Under the conditions $0\le c_D\le{\cal D}$ and $0\le c_E\le{\cal E}$, the above equations have one and only one fixed point. This point is always stable and, asymptotically, the cytosolic distributions will approach the corresponding stationary values $C_D$ and $C_E$, respectively. In this limit, the dynamics of the Min proteins is described by two partial differential equations for the distributions of the proteins bound to the membrane
\begin{eqnarray}
\label{eq:dynHomD}
\partial_t c_d & = & \omega_DC_D(c_{\rm max}-c_d-c_{de}) - \omega_{E}C_Ec_d-\partial_xj_d\\
\label{eq:dynHomDE}
\partial_t c_{de} & = & -\omega_{de}c_{de} + \omega_EC_Ec_d\quad.
\end{eqnarray}
Note, that the reaction terms in these equations are linear and describe relaxation to a stationary value; only the current contains non-linearities and can generate an instability. This feature distinguishes this system from classical reaction-diffusion systems, where transport is due to diffusion and where instabilities are created by the reaction terms.

\begin{figure}
\centering
\includegraphics*[width=12 truecm, angle=0]{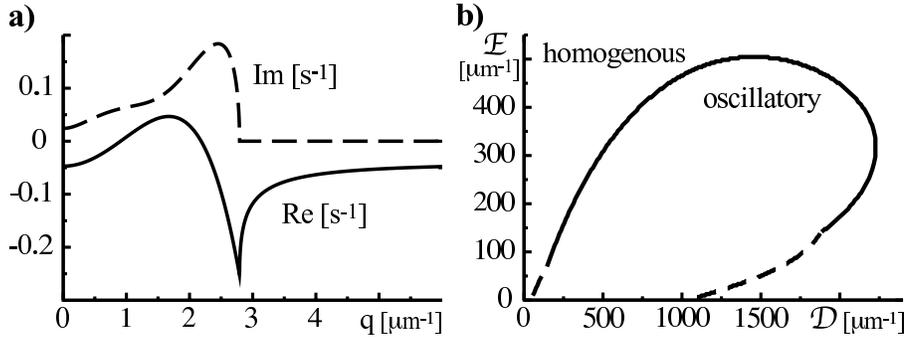}
\caption{
Linear stability of the homogenous state. 
a) Real (Re, solid line) and imaginary part (Im, dashed line) of the eigenvalues of the linear operator describing the dynamics of small perturbations around the homogenous state as a function of the wave number $q$. Modes with wave numbers between 1$\mu$m$^{-1}$ and 2.2$\mu$m$^{-1}$ are oscillatory and unstable. 
b) Stability of the homogenous state as a function of the average total MinD and MinE densities ${\cal D}$ and ${\cal E}$. The solid line indicates a line of oscillatory instabilities while the dashed lines indicate stationary instabilities. 
The values of the parameters are $\omega_{D}=4\cdot10^{-5}\mu $m$s^{-1}$, $\omega_{E}=3\cdot10^{-4}\mu $m$s^{-1}$, $\omega_{de}=0.04s^{-1}$, $D_{d}=0.06\mu $m$^{2}s^{-1}$, $c_{\rm max}=1000\mu $m$^{-1}$,  $k_{1}=1.5 \cdot10^{-6}\mu $m$^{4}s^{-1}$, $k_{2}= 1.8\cdot10^{-7}\mu $m$^{6}s^{-1}$, $\bar k_{1}=-1.2\cdot10^{-6}\mu $m$^{4}s^{-1}$, $\bar k_{2}=1.2\cdot10^{-10}\mu $m$^{6}s^{-1}$. In (a) ${\cal D}=900\mu $m$^{-1}$ and ${\cal E}=350\mu $m$^{-1}$.
}
\label{fig:linstab}
\end{figure}

The homogenous state $c_{d}(x) = {\cal D}-{\cal E}-C_D+C_E$ and $c_{de}(x)={\cal E}-C_E$ is a stationary state of the dynamic equations (\ref{eq:dynHomD}) and (\ref{eq:dynHomDE}). It is stable, unless $k_{1}$ exceeds a critical value $k_{1,c}$.
The results of a linear stability analysis for a supercritical value of
$k_{1}$ are shown in Fig.~\ref{fig:linstab}a.
The stability region of the homogenous state as a function of the total MinD and MinE concentrations, $\cal D$ and $\cal E$, is shown in Fig.~\ref{fig:linstab}b. At the instability an inhomogenous stationary state appears if the detachment rate of MinDE complexes from the membrane is above a certain critical value, $\omega_{de}>\omega_{de,c}$.  In the opposite case, an oscillatory state appears. 
Oscillatory instabilities only occur if the protein density on
the membrane cannot exceed a maximal value $c_{\rm max}$. For an oscillatory instability 
the unstable mode is of the form
\begin{eqnarray}
c_{d} & \propto &\cos(\Omega_{c}t)\cos(q_{c}x)\\
c_{de} & \propto & \cos(\Omega_{c}t+\phi)\cos(q_{c}x)
\end{eqnarray}
This standing wave reflects the qualitative features of the observed Min-oscillations. The wave number $q_{c}=n\pi/L$, where $n$ is a natural number, and the frequency $\Omega_{c}$ of the critical mode depend on the system parameters. For instance, we find 
\begin{equation}
q_{c}^{4} =
\frac{(\omega_{D}C_{D}+\omega_{de}+\omega_{E}C_{E})}{C_{d}(c_{\rm max}-C_{d}-C_{de})k_{2}}\quad,
\end{equation}
and if $\bar k_{1}=\bar k_{2}=0$
\begin{eqnarray}
\Omega_{c}^{2} & = & \omega_D \omega_E 
C_D C_E-\omega_{de}^{2}
\end{eqnarray}

\begin{figure}
\centering
\includegraphics*[width=12 truecm, angle=0]{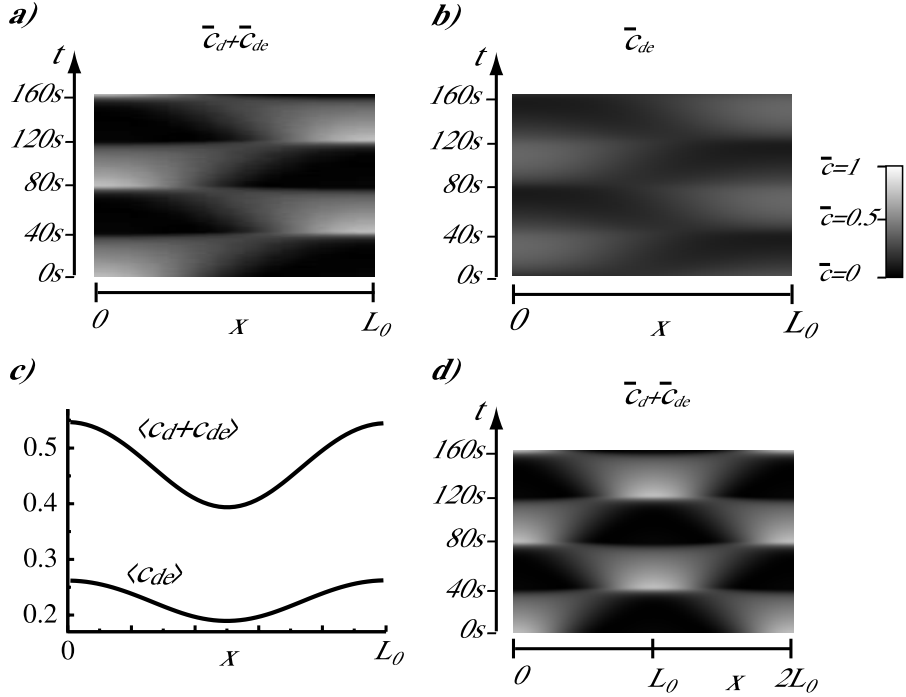}
\caption{
Oscillatory solutions of the dynamic equations (\ref{eq:dynHomD}) and (\ref{eq:dynHomDE}).
a,b) Space-time plots of the total MinD and MinDE distributions on the membrane, $\bar{c}_{d}+\bar{c}_{de}=(c_{d}+c_{de})/c_{\rm max}$ and  $\bar{c}_{de}=c_{de}/c_{\rm max}$, respectively, for system size $L_{0}=2\mu$m. Both distributions show pole-to-pole oscillations with a temporal period of about 80s. 
c) The total MinD and the MinDE distribution averaged over one temporal period shown in (a) and (b), $\left\langle \bar c_{d}+\bar c_{de}\right\rangle$ and $\left\langle \bar c_{de}\right\rangle$. Both distribution display a clear minimum at $x=L_{0}/2$. 
d) Space-time plot of the total MinD distribution on the membrane, $\bar{c}_{d}+\bar{c}_{de}$ for system size $2L_{0}$. The pattern has doubled as compared to the pattern in the system of length $L_{0}$. Parameters are $k_{1}=2.1 \cdot10^{-6}\mu m^{4}s^{-1}$, $k_{2}= 2.5\cdot10^{-7}\mu m^{6}s^{-1}$, and the remaining values as in Fig.~\ref{fig:linstab}a.
}
\label{fig:osc}
\end{figure}

\begin{figure}
\centering
\includegraphics*[width=12 truecm, angle=0]{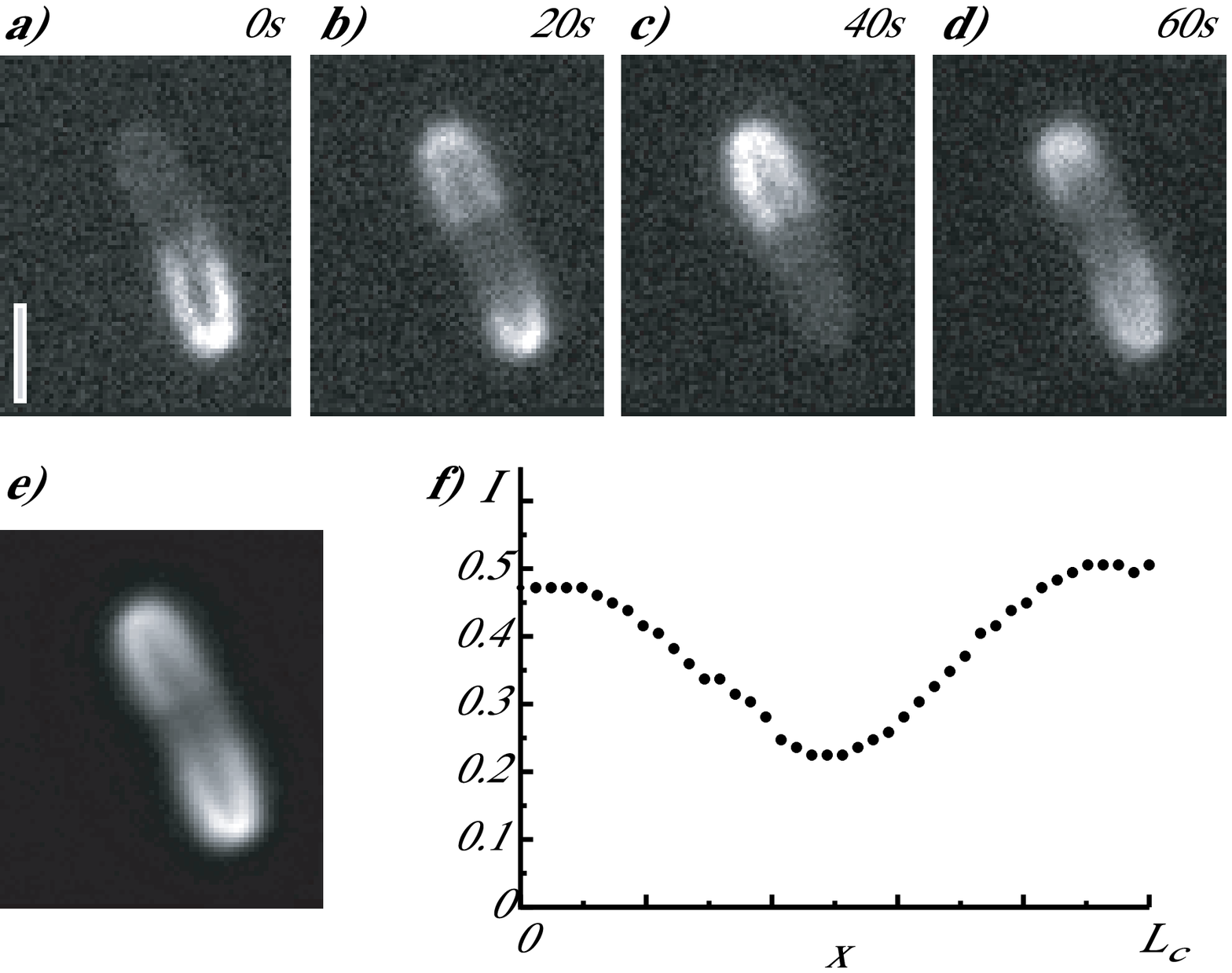}
\caption{
Oscillations of MinD-GFP in \textit{E.~coli}. a-d) Fluorescence images of MinD-GFP in a cell at subsequent time points separated by 20s. e) Time-average of all frames during one oscillation period. Two subsequent frames are separated by 1s. f) Fluorescence intensity I obtained from a line scan of the fluorescence signal in (e). The background signal has been subtracted from the total signal which has then been rescaled with the maximum intensity during the oscillation. The slight asymmetry is due to bleaching during the observation period. Scale bar: 1$\mu$m. The cell length is $L_{c}=2.3\mu$m.}
\label{fig:oscExp}
\end{figure}

The oscillatory patterns can be obtained from numerical integration of the dynamic equations (\ref{eq:dynHomD}) and (\ref{eq:dynHomDE}). A typical example is shown in Fig.~\ref{fig:osc}a, b. For some time the total MinD-distribution $c_{d}+c_{de}$ is localized in one half and then switches to the other. In this process, the transition time is very short as compared to the dwell time in one half. The MinE distribution shows a similar behavior, but the transition between the two halves is less rapid. The time-averaged distribution of both, MinD and MinE shows a minimum in the center and increases towards the system boundaries, see Fig.~\ref{fig:osc}c. The parameters have been chosen such that the temporal period is about 80s, which is similar to the values observed in experiments with fluorescently labeled MinD, see Fig.~\ref{fig:oscExp}. The figure also displays the time-averaged MinD-distribution  
with a minimum in the center. 
In the case displayed on Fig.~\ref{fig:oscExp}f, the minimum at the center is more pronounced
than for the theoretical calculation: while experimentally the minimum
is at about 50\% of the maximum, it is at about 70\% in the numerics. This might indicate the need for further non-linearities in the theory. 
However, for other cells examined, the minimum 
is much shallower or even absent  (data not shown). This might reflect
deviations in the total protein density in individual bacteria from the average total protein density in a bacterial colony. Note also, that in
the numerics, the value of the minimum decreases with 
the system length
up to the point the oscillation pattern acquires a new stripe. It would be 
interesting to experimentally test this dependence of the average MinD
distribution on the cell length. Due to fast bleaching of the GFP we
were not able to perform this experiment.

In the model, the transition of MinD from one half to the other can be understood as follows. If MinD is localized in one half, MinE will bind and drive MinD off the membrane. Although the distribution of cytosolic MinD is homogenous, MinD preferentially binds in the other half, because there are more available binding sites. The resulting inhomogeneity of membrane-bound MinD is then amplified by MinD aggregation. As a consequence of the homogenous distribution of cytosolic MinE, the spatial dependence of the attachment rate of MinE follows the profile of membrane-bound MinD, and the distribution of MinDE complexes is similar to the one of MinD on the membrane, see Fig.~\ref{fig:osc}a, b. In particular, the positions of the maxima of $c_{de}$ are linked to the position of the maxima of $c_{d}$. In the example given in Fig.~\ref{fig:osc}a, b, maxima are always located at the boundaries $x=0$ and $x=L$. 

\begin{figure}
\centering
\includegraphics*[width=12 truecm, angle=0]{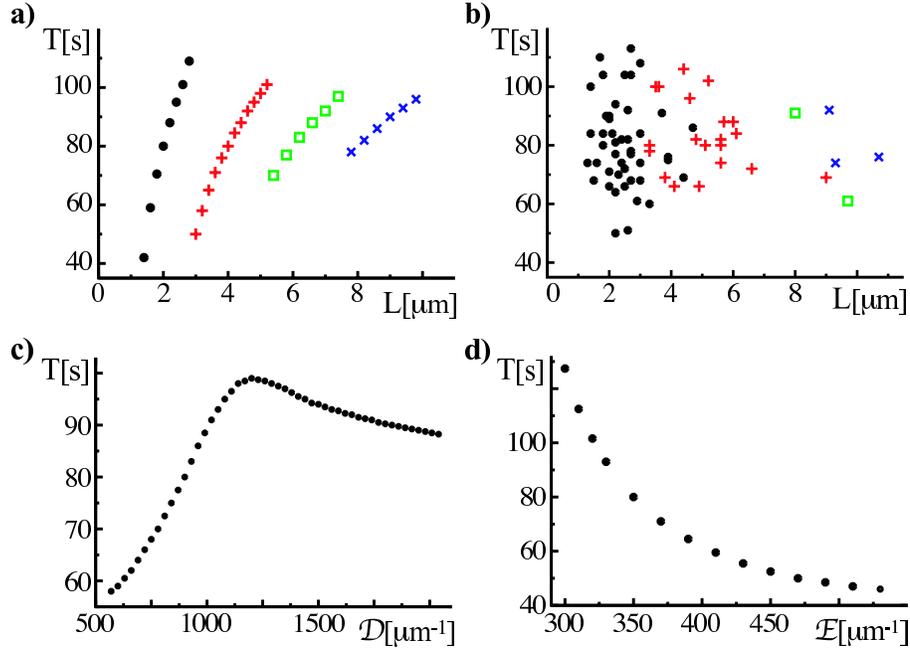}
\caption{
Dependence of the oscillation period $T$ on parameters. a) Oscillation period of solutions to the Eqs.~(\ref{eq:dynHomD}) and (\ref{eq:dynHomDE}) as a function of the system length. Black dots: oscillation pattern as in Fig.~\ref{fig:osc}a, red crosses: oscillation pattern as in Fig.~\ref{fig:osc}b, green and blue dots: oscillation pattern with three and four stripes, respectively. For the system length where the oscillation pattern changes, the period shows a discontinuity. b) Oscillation period measured for \textit{E.~coli}. The different symbols refer to the same oscillation patterns as in (a), error bars are of about the size of the symbols. c) Oscillation period in the model as a function of the average total MinD concentration $\cal D$. The period initially increases and then decreases slightly with $\cal D$. d) The same as (c) but for the average total MinE concentration $\cal E$. The period decreases with the amount of MinE.
For (a), (c), and (d), the parameters values are as in Fig.~\ref{fig:osc}a, the system length in (c) and (d) is $2\mu$m.
} \label{fig:TvsLjumps}
\end{figure}

As the system size is increased, the patterns change and striped patterns for $c_{d}$ and $c_{de}$ appear, see Fig.~\ref{fig:osc}d. This reflects the finite wave number of the critical mode. In addition to changes in the oscillation pattern, the temporal period, too, changes as the system size is varied. It increases monotonically with the system size, but at certain sizes jumps back towards a lower value, see Fig.~\ref{fig:TvsLjumps}a. The discontinuities occur for the system sizes where the oscillatory pattern acquires a new ``stripe''. For the parameter values used here, a new stripe appears for a system size  of 3$\mu$m. 

We measured the temporal period of the oscillations in \textit{E.~coli} containing MinD-GFP, Fig.~\ref{fig:TvsLjumps}b (see Methods). The periods fall in the range of 50s to 120s, even for bacteria of 10$\mu$m in length. 
The data indicate large variations of the oscillation period for
cells of approximately the same length. This might be due to differences
in the MinD and MinE concentrations for different bacteria and thus
reflect individuality of the cells. An experimental verification would
require the measurement of the protein concentration in an individual
cell together with the temporal period of the Min oscillations.
The  oscillation periods found for the dynamic equations
(\ref{eq:dynHomD}) and (\ref{eq:dynHomDE}) span the same range
as the experimentally observed. Furthermore, experimentally we
 observed striped oscillation patterns only for bacteria longer than
3$\mu$m, however, there is no sharp transition length for which
the pattern changes. This behaviour, too, could be due to 
variations in the protein densities  between different bacteria.

In the model, the temporal oscillation period also depends on the total MinD and MinE concentrations, $\mathcal{D}$ and $\mathcal{E}$, see Fig.~\ref{fig:TvsLjumps}c, d. It increases monotonically with the amount of MinD until it starts to descend slightly. As a function of the number of MinE molecules, the period decreases. Both dependencies are compatible with experimental observations where the 
period has been found to increase with the MinD concentration and to decrease with the MinE concentration~\cite{RdeB99}, but only few data points have been reported and further measurements are necessary. 

In conclusion, the solutions to the dynamic equations presented here
are compatible with the experimental data, but further experiments 
are necessary in order to verify the discontinuous dependence of
the oscillation period on the system length.

\section{The MinE-ring}

Early experiments indicated an accumulation of MinE close to the cell center~\cite{rask97}. This accumulation is commonly referred to as the MinE-ring.  More recently, using deconvolution techniques, MinE has been found to be arranged in a helix with accumulation close to the cell center and, although weaker, at the cell poles~\cite{SLR03}. It has been suggested that the helical arrangement of MinE is induced by the helical arrangement of MinD and that the accumulation of MinE occurs at the ends of the MinD helix~\cite{SLR03}. In cells mutant for MinE, oscillations have been observed in the absence of a MinE-ring~\cite{RFSZCR00}. In that case, the temporal period is larger than in non-mutant cells. Still, this experiment clearly shows that the MinE-ring is not necessary for the oscillations. 

In the one-dimensional description presented above, MinE-rings correspond to maxima in the MinDE distribution. In the examples given so far, such maxima only occur at the system boundaries. For system lengths close to the value at which the pattern acquires a new stripe, maxima can be detected closer to the system's center. However,
this is unlikely to be the mechanism for MinE-ring formation in \textit{E.~coli},
because no dependence of the existence of the ring on the cell size has been reported. Furthermore, as argued above, in the limit of homogenous cytosolic MinD- and MinE-distributions, maxima in the MinDE-distribution are induced by maxima in the MinD-distribution.  MinD-rings are not observed experimentally, though.

There are at least three other possible mechanisms that can in principle account for the observed accumulation of MinE at the ends of the MinD-helix. In the first mechanism, the diffusion length of cytosolic MinE, $l_{E}=(D_{E}/\omega_{E}c_{\rm max})^{1/2}$ , is shorter than half of the cell length. In this case, cytosolic MinE will predominantly attach before it has reached the opposite cell pole, which might lead to an accumulation close to the cell center. This seems to be the mechanism of MinE-ring formation in the models proposed in Refs.~\cite{HRdeV01,HMW03}. Accordingly, the ring vanished in \cite{HMW03} when the attachment rate of MinE was reduced, leading to an increase of $l_{E}$. To test whether this mechanism is supported by the Eqs.~(\ref{eq:dynD})-(\ref{eq:dynde}), we studied the system for finite values of $D_{D}$ and $D_{E}$. In this case the cytosolic distributions $c_{D}$ and $c_{E}$ are not homogenous and all four equations have to be solved simultaneously. For the parameter values considered above, the oscillation patterns do not change significantly as long as the diffusion constants $D_{D}$ and $D_{E}$ are larger than 0.1$\mu$m$^{2}/s$ and no maxima of $c_{de}$ independent of maxima of $c_{d}$ were found. The diffusion length $l_{E}$ is also influenced by the value of $\omega_{E}$. For $D_{D}=D_{E}=2.5\mu$m$^{2}/s$, the values expected for diffusion in the cytosol, and values of $\omega_{E}$ smaller than $3.2\cdot 10^{-4}\mu$m/s the same behavior was found. Still larger values of $\omega_{E}$ destroy the oscillations. 
Note, that by assumption, the one-dimensional description is only appropriate if the diffusion length $l_{E}$ is larger than the cell diameter, i.e., $l_{E}\ge 1\mu$m. We conclude, that this mechanism is not supported by the dynamic equations presented above and can be tested only in a three-dimensional description.

Two other mechanisms of minE-ring formation are suggested by studies of kinesin-subfamily Kin13 members~\cite{desa99,hunt03}. These proteins induce the depolymerization of microtubules. In this process they accumulate at both ends of the microtubule. As MinE might act on MinD filaments in much the same way, accumulation of MinE could follow from a similar mechanism as accumulation of the Kin13-kinesins. The latter could be a consequence of a higher affinity of the microtubule end for binding the motor. Related ideas for the binding of MinE to MinD have been proposed in \cite{MdeB01} and also in \cite{K02}. The analogy with Kin13-kinesins offers still another explanation for the accumluation of MinE, namely dynamic accumulation due to processive depolymerization \cite{KKCJ04}. The present framework for studying the dynamics of Min-proteins is not suited for studying these effects, as filaments are not explicitly incorporated. Work in this direction is in progress.

\section{Conclusion and outlook}

We have presented a phenomenological description of the dynamics
of MinD and MinE in \textit{E.~coli}. The description is based on 
the binding of MinD to the cytosolic membrane, recruitment of 
MinE to the membrane by membrane-bound MinD, MinE-induced detachment of MinD,
as well as an interaction between molecules bound to the 
membrane. 
For a sufficiently strong  attraction between membrane-bound MinD-molecules, 
these processes generate pole-to-pole oscillations of the Min-proteins.   
The phenomenological form of the current for membrane-bound MinD used in the present work captures generic features of the protein interaction and does not refer to a specific microscopic mechanism. It allows for a quantitative comparison between the oscillatory solutions of the dynamic equations and experimental findings. In agreement with the latter, oscillations with a temporal period from 40s to 120s can be obtained. This value is essentially determined by the detachment rate $\omega_{de}$ of MinDE-complexes. For the parameter values given in the text, the oscillatory pattern acquires a second stripe for a system size of 3$\mu$m, which agrees well with the smallest bacterial length
for which period doubling is seen in Fig.~\ref{fig:TvsLjumps}b. This length is essentially determined by the ratio of the parameters $k_{1}$
and $k_{2}$.  Finally, the time-averaged MinD-distribution shows a
minimum at the cell center as we  observed experimentally. Starting
from  an almost homogenous average distribution, the 
depth of the minimum increases with the system length. In \textit{E.~coli}, this 
feature could obviously be used to couple assembly of the
Z-ring to the cell length and hence to control the cell cycle.

The phenomenological description of the aggregation current can be related to microscopic descriptions
of the protein dynamics. A simple process leading to aggregation
is based on short-range pair interaction
potentials. In this case, the phenomenological parameters are linked to microscopic quantities by~\cite{ME95}
\begin{eqnarray}
\label{eq:k1micro}
k_1 & = & \frac{1}{c_{\rm max}^{2}}\frac{D_d}{k_BT}U,\\
\label{eq:k2micro}
k_2 & = & \frac{1}{c_{\rm max}^{2}}\frac{D_d}{k_BT}U r^2
\end{eqnarray}
and analogously for $\bar k_{1}$ and $\bar k_{2}$. Here, $U$ measures the strength of the MinD-interaction potential, $r$ is a typical length scale of the interaction, and $k_BT$ thermal energy. These relations are valid whenever $r$ is much smaller than the diffusion length $l_{d}=\sqrt{D_{d}/\omega_{E}c_{\rm max}}$. Assuming, as
done above,
a diffusion constant of 0.06$\mu$m$^{2}/s$ for membrane bound
MinD, which falls well in the regime of measured diffusion constants 
for membrane proteins~\cite{lipp01},
the values of the phenomenological coefficients used above
imply values of 35$k_{B}T$ for the interaction strength between membrane-bound MinD and 20$k_{B}T$ between MinD and MinDE complexes. The
range for MinD-MinD interactions is then 350nm and for MinD-MinDE
interactions 10nm. While all other values are acceptable, the
range for MinD-MinD interactions is too large for pure electrostatic 
interactions. This points to
more involved microscopic dynamics of membrane-bound MinD  
than discussed here.

Our analysis of the dynamic equations (\ref{eq:dynD})-(\ref{eq:dynde}) has focused on the case of homogenous cytosolic distributions of MinD and MinE, $c_{D}$ and $c_{E}$. Solutions in this limit are very similar to solutions to the full equations if the diffusion constants of both MinD and MinE have the realistic value of 2.5$\mu$m$^{2}$/s. This implies that the approximation of constant $c_{D}$ and $c_{E}$ is appropriate. Apart from providing a reduced set of equations that is more convenient to study than the four equations of the full system, this approximation might also have an important implication regarding experiments. One might expect that oscillations should be observable in a purified system containing essentially only MinD, MinE, and phospholipid vesicles. The analysis presented here suggests that oscillations will show up in presence of a homogenous distribution of cytosolic proteins. Therefore, the closed geometry of the bacterium might not be essential and an open geometry could be used instead.  A second implication of our analysis is that the number of available binding sites might need to be limited in order to produce oscillations. 

Other mechanisms that have been suggested for the Min-oscillations agree in the essential assumptions with the one studied here, namely the ability of ATP-dependent binding of MinD to the membrane, the recruitment of MinE to the membrane by MinD, and the release of MinD from the membrane driven by MinE. The proposed mechanisms differ, however, in essential points. Meinhardt and deBoer suggested that protein synthesis might be an essential element~\cite{MdeB01}, which is not supported by experiments where the synthesis of proteins was interrupted and the oscillations still continued~\cite{RdeB99}. Howard et al. assume that MinD and MinE form complexes in the cytosol and bind together to the membrane~\cite{HRdeV01}. This implies in particular an exponential increase of the temporal period of the oscillations with the system length, with a period of 1000s for
a system of length 7$\mu$m. This is qualitatively different from the behavior reported for the mechanism studied in this work, 
see Fig.~\ref{fig:TvsLjumps}a. The experimental data presented in   Fig.~\ref{fig:TvsLjumps}b show oscillation periods that do not exceed
120s for bacteria of a length up to 10$\mu$m. However, more
experiments are needed, in particular to obtain simultaneously
values for the protein densities and the oscillation period of individual
bacteria.

The system studied by Huang et al. differs from the one studied here in the way MinD-aggregates are formed on the membrane~\cite{HMW03}. There MinD aggregation follows a one-step process: attachment to the membrane occurs with a higher rate at locations where MinD is already bound. In contrast we considered a two step-process, namely, cytosolic MinD binds first to the membrane and only then self-assembles into a filament. This difference might at first sight seem minor. However, it leads to striking differences in the model behaviours. First of all, assuming a one-step process for MinD aggregation, a three-dimensional geometry as well as a finite ATP-exchange rate is required to generate striped oscillation patterns in long systems. Secondly, in the model by Huang et al. there are no oscillatory solutions at all for homogenous cytosolic distributions. Furthermore, as discussed above, in the model studied here, MinE-rings have not been found to form by the mechanism underlying formation of the MinE-ring in the model by Huang et al. The differences in the mechanism for MinD aggregation thus lead to striking consequences for the collective behaviour of the Min proteins. One possibility to discriminate between the two mechanisms is by studying the dynamics of Min proteins that are not confined to a cell. Further analysis of the models might lead to other possible key experiments.

In combination, all proposed mechanisms underlying the Min oscillations suggest new experiments that will allow us to understand the Min-oscillations better. In order to make even closer contact with experiments, the formation of MinD helices must be included. Fluctuations due to the moderate number of Min-molecules might play an important role. First attempts in studying the influence of fluctuations on the oscillations have been undertaken~\cite{HR02,KHKW04}, but further work is needed and will probably yield results of relevance for pattern formation in the presence of noise beyond the Min-system.

\section{Glossary}

\begin{description}
\item[Mini-cell.] DNA-free small cell that is produced by \textit{E.~coli} dividing close to a cell pole.
\item[Min proteins.] Proteins involved in the determination of the
division site. Mutations in these proteins lead to the formation of
mini-cells.
\item[Linear stability analysis.] In a linear stability analysis the stability 
of a stationary state against small perturbations is assessed by 
linearizing the dynamic equations with respect to the stationary state.
\item[Reaction diffusion system.] Several
reacting substances that are transported in space through diffusion.
The reactions can induce instabilities of a stationary homogenous
distribution leading to the formation of spatio-temporal patterns. 
\item[Kinetic hopping model.] Particles are confined to the
sites of a lattice. Motion of the particles is described by hopping 
between sites.
\end{description}

\ack

We thank M.~B\"ar, R.~Everaers, F.~J\"ulicher, H.~Chat\'e,  A.~Politi, and J.~Lutkenhaus for valuable discussions, M.~De Menech for help with the figures, and S.~Diez, R.~Hartmann, I.~Riedel, and J.~Howard for support with the experiments.

\appendix
\section{Dynamics in three dimensions and reduction to one dimension}
\label{app:reduction}

In this Appendix it is shown, how the dynamics in three spatial dimensions can
effectively be reduced to a description in one spatial dimension. The bacterium 
is conveniently approximated by a cylinder with radius $R_0$ and length $L$. 
The volume densities of cytosolic MinD and MinE at a given point are $c_D(r,\vartheta,x)$ 
and $c_E(r,\vartheta,x)$, respectively. Here, $r$ and $\vartheta$ denote the radial 
and azimuthal coordinate, respectively, while $x$ is the coordinate along the long axis. 
Their time evolution is goverened by
\begin{eqnarray}
\partial_t c_D(r,\vartheta,x)& = -\omega_D(c_{\rm max}-c_d(\vartheta,x)-c_{de}(\vartheta,x)) 
c_D(r,\vartheta,x)\delta(r-R_0)\nonumber\\
&+\omega_{de}c_{de}(\vartheta,x)\delta(r-R_0)
+D_D \Delta_{3d} c_D(r,\vartheta,x)\\
\partial_t c_E(r,\vartheta,x)& =  -\omega_{E}c_d(\vartheta,x)c_E(r,\vartheta,x)\delta(r-R_0)+\omega_{de}
c_{de}(\vartheta,x)\delta(r-R_0)\nonumber\\ 
&+D_E\Delta_{3d} c_E(r,\vartheta,x)\quad.
\end{eqnarray}
Here, $c_d$ and $c_{de}$ are the surface densities of membrane-bound MinD and MinDE-complexes, 
$\Delta_{3d}$ is the three-dimensional Laplace-operator and the factors
of $\delta(r-R_0)$ restrict attachment to and detachment from the cytoplasmic membrane 
to a region adjacent to the cell wall.

Since the diffusion constant of cytosolic MinD and MinE is of the order of $1\frac{\mu{\rm m}^2}{s}$, 
whereas the period of the oscillations is about 1min, it is reasonable to assume the density of 
cytosolic MinD and MinE to be homogenous perpendicular to the bacterial long axis. 
The volume densities of cytosolic MinD and MinE can then be replaced by surface densities 
$\tilde c_D$ and $\tilde c_E$  with
\begin{eqnarray}
c_D(r,\vartheta, x) & = & \frac{1}{R_0}\tilde c_D(\vartheta,x)\\
c_E(r,\vartheta, x) & = & \frac{1}{R_0}\tilde c_E(\vartheta,x)\quad.
\end{eqnarray}
Then, the equations governing the evolution of the protein densities read
\begin{eqnarray}
\partial_t \tilde c_D & = & -\frac{\omega_D}{R_0}(c_{\rm max}- c_d-c_{de})\tilde c_D 
+ \omega_{de}c_{de}+D_D\Delta_{2d}\tilde c_D\\
\partial_t \tilde c_E & = & -\frac{\omega_E}{R_0}c_d\tilde c_E+\omega_{de}c_{de}+D_E\Delta_{2d}\tilde c_E\\
\partial_t c_d & = & \frac{\omega_D}{R_0}(c_{\rm max}- c_d-c_{de})\tilde c_D - 
\frac{\omega_E}{R_0}c_d\tilde c_E -\nabla\cdot {\bf j}_{d}\\
\partial_t c_{de} & = & \frac{\omega_E}{R_0}c_d\tilde c_E-\omega_{de}c_{de}\quad,
\end{eqnarray}
where ${\bf j}$ is the aggregation current of MinD on the inner cell membrane and $\Delta_{2d}$ 
the two-dimensional Laplace operator on the cylinder surface.

It has been shown that MinD forms a filamentous structure on the inner cell membrane~\cite{SLR03}. 
Projection on this structure  yields line-densities, e.g., 
$\bar c_d(x)=\int_0^{2\pi} c_d(\vartheta,x)R_0\;d\vartheta$. They are connected to the surface densities via
\begin{eqnarray}
\tilde c_D(\vartheta,x)\approx\frac{1}{2\pi R_0}\bar c_D(x)\\
\tilde c_E(\vartheta,x)\approx\frac{1}{2\pi R_0}\bar c_E(x)\\
c_d(\vartheta,x) \approx \bar c_d(x)\delta(\vartheta-\vartheta(x))\\
c_{de}(\vartheta,x)\approx \bar c_{de}(x)\delta(\vartheta-\vartheta(x))\quad,
\end{eqnarray}
where $\vartheta(x)$ parametrizes the MinD-helix on the inner cell membrane. 
The dynamic equations for the line densities $\bar c_D$, $\bar c_E$, $\bar c_d$, 
and $\bar c_{de}$ are then given by Eqs.~(\ref{eq:dynD})-(\ref{eq:dynde}).
The current ${j}_{d}$ appearing there is obtained by projection of
the surface current ${\bf j}_{d}$ on the $x$-direction. 
Note, that a description of the formation of MinD-helices would require 
also a specification of the perpendicular component of the current
${\bf j}_{d}$.

\section{Methods}

Bacteria of the \textit{E.~coli} K12 strain JS964 were generously donated by J.~Lutkenhaus, University of Kansas. Bacteria taken from the freezer were grown overnight in 3ml Luria-Bertani (LB) medium at 37$^{\circ}$C together with 3$\mu$l spectinomycin. Of the overnight culture 500$\mu$l together with 50$\mu$l spectinomycin were given in 50ml LB medium and grown for two hours at 37$^{\circ}$C. Expression of MinD-GFP was induced by 50$\mu$l IPTG and growing the bacteria at 31$^{\circ}$C for at least one hour. Bacteria were immobilized for fluorescence imagery by using silane-coated cover slips. Fluorescent images were taken at room temperature with an inverted microscope (Axiovert 200M, Zeiss) using a CCD camera from Spot Diagnostic Instruments, Inc.~driven by Metavue, Universal Imaging. The frame
rate for measuring the time-average in Fig.~\ref{fig:oscExp} was 1Hz
and varied between 0.33Hz and 1Hz for the data in Fig.~\ref{fig:TvsLjumps}b. Data were analysed using Metamorph, Universal Imaging.

\end{document}